\documentclass[doublecol]{epl2}

\newcommand{\be}{\begin{equation}}\newcommand{\ee}{\end{equation}}
\newcommand{\bea}{\begin{eqnarray}}\newcommand{\eea}{\end{eqnarray}}
\newcommand{\brr}{\begin{array}}\newcommand{\err}{\end{array}}
\newcommand{\bit}{\begin{itemize}}\newcommand{\eit}{\end{itemize}}
\newcommand{\ben}{\begin{enumerate}}\newcommand{\een}{\end{enumerate}}

\newcommand{\ba}{\begin{array}}
\newcommand{\ea}{\end{array}}

\def\lan{\langle}

\def\ran{\rangle}

\def\1{{_{1}}}\def\2{{_{2}}}

\def\noHe0{:\;\!\!\;\!\!:H_e(0):\;\!\!\;\!\!:}
\def\noHm0{:\;\!\!\;\!\!:H_\mu(0):\;\!\!\;\!\!:}

\title{Flavor entanglement in neutrino oscillations in the wave packet description}
\shorttitle{Flavor entanglement in neutrino oscillations} 

\author{Massimo Blasone\inst{1,2} \and Fabio Dell'Anno\inst{2,3,4} \and Silvio De Siena\inst{2,3,5}
\and Fabrizio Illuminati \inst{2,3,5}}
\shortauthor{M.~Blasone, F.~Dell'Anno, S.~De~Siena and F.~Illuminati}

\institute{
  \inst{1} Dipartimento di Fisica, Universit\`a di Salerno, Via Giovanni Paolo II, 132
84084 Fisciano, Italy
\\
  \inst{2} INFN Sezione di Napoli, Gruppo collegato di Salerno, Italy
\\
 \inst{3} Dipartimento di Ingegneria Industriale, Universit\`a   di Salerno, Via Giovanni Paolo II, 132
84084 Fisciano, Italy
\\
 \inst{4}
 Liceo Statale P.E. Imbriani, Via Pescatori 155, I-83100 Avellino, Italy
\\
 \inst{5}
CNISM  Unit\`a di Salerno, I-84084 Fisciano (SA), Italy

}


\abstract{
The wave packet approach to neutrino oscillations provides an enlightening description of quantum decoherence
induced, during propagation, by localization effects.
Within this approach, we show that a deeper insight into the dynamical aspects
of particle mixing can be obtained if one investigates the behavior of quantum correlations associated to flavor oscillations.
By identifying the neutrino three-flavor modes with (suitably defined) three-qubit modes,
the exploitation of tools of quantum information theory for mixed states allows
a detailed analysis of the dynamical behavior of flavor entanglement during free propagation.
This provides further elements leading to a more complete understanding of
the phenomenon of neutrino oscillations, and  a basis for possible applicative implementations.
The analysis is carried out by studying the distribution of the flavor entanglement;
 to this aim, we perform combined investigations of the behaviors of the two-flavor concurrence and of the logarithmic negativities associated with specific bipartitions of the three flavors.
}

\begin{document}

\maketitle

\section{Introduction}
The study of entanglement in elementary particle systems belongs to the current frontiers of research in Physics~\cite{Bert1,Bert2,DiDomenico:2011fq,NoiPRD,NoiEPL,Blasone:2013yoa,Kayser10,Smirnov11,Boyanovsky:2011xq,Lello:2013bva,NoiEPL2,NoiHindawi,EPLBrazil,Bernardini:2012uf}. In fact, the study of quantum correlations in these systems can offer a complementary
and  enlightening  viewpoint  for  understanding  fundamental phenomena.
Furthermore, it represents a prerequisite for a subsequent analysis in applicative contexts
concerning, for instance, the implementation of quantum information protocols in the framework
of elementary particle physics and quantum field theory.

The phenomenon of flavor oscillations associated with the neutrino mixing must be considered
a significant instance of dynamical entangled system
characterized by single-particle, multi-mode entanglement~\cite{NoiPRD,NoiEPL,NoiEPL2,NoiHindawi,EPLBrazil,Alok:2014gya}.
By using proper entanglement measures, it has been shown that flavor entanglement can be written
in terms of the flavor transition probabilities;
furthermore, it has been also
proposed an experimental scheme that allows the transfer
to spatially delocalized two-flavor charged lepton states of the quantum information encoded in neutrino states~\cite{NoiEPL}.
The analysis inherent the quantification and characterization of multi-mode flavor entanglement
in the oscillating neutrino system has been carried out both in a quantum-mechanical setting~\cite{NoiPRD,NoiEPL,EPLBrazil}
and in the framework of quantum field theory~\cite{NoiEPL2,NoiHindawi}, while decoherence effects have been investigated in Refs.~\cite{NoiPRD,EPLBrazil}.
In these articles, the multipartite entanglement in the dynamics of flavor oscillations was analyzed by using, as reference space,
the (three-qubit) Hilbert space associated with neutrino mass eigenstates.
By adopting the wave-packet description for the  mass  eigenstates,  it  was  shown  a  strict  connection
between the decoherence effects and the spatial behavior of quantum entanglement; in fact, due to the differences
among the neutrino masses, the corresponding wave packets propagate at different group velocities, thus resulting
in a mutual spatial separation that increases during the propagation.  Consequently, the free evolution leads to a
progressive loss of the coherent interference effects, that are connected with the destruction of the oscillation
phenomenon and with the vanishing of the multipartite quantum entanglement among mass modes.

On the other hand, the (anyway interesting) approach based on the mass eigenstates involves some drawbacks,
since it deals with a not directly observable context.
In the present paper we accomplish a further step forward in the analysis of the dynamic of quantum
correlations by considering the propagation of flavor neutrino states, which directly enter into the production and
detection processes.

By using the wave packet approach, we analyze the behavior of multipartite entanglement relative to the
(three-qubit) Hilbert space of flavor neutrino eigenstates.
{Although the neutrino system is described in principle by a pure state, we obtain a mixed state after a time integration has been performed, following a standard procedure. Consequently, in order to  quantify the entanglement content of such a mixed state,  we exploit two measures, the concurrence and the logarithmic negativity, that were specifically devised for this case.
In detail, we consider the content of entanglement shared by two given flavors after  a partial trace has been performed with
respect to the third flavor (concurrence), and the content of entanglement in bipartitions of the three-flavor system
(logarithmic negativity).}
We show that the transition probabilities provide only partial information about the dynamics of quantum correlations,
and that the use of quantum information methods leads to a deeper insight.
Indeed, such correlations are not fully encoded into flavor oscillation transition probabilities, but can be unveiled by
means of appropriate experimental protocols. In particular, it is worth to be remarked that the combined exploitation of the two, operationally different, measures provides indications about the distribution of the entanglement among the different flavors.

The paper is organized as follows.  First we review the wave-packet approach for free propagating neutrinos, and
the main aspects related to flavor oscillations.  After introducing the suitable entanglement measures, we move
to the results by investigating the behavior of the entanglement associated with flavor oscillations.
Finally, we draw our conclusions.


\section{Neutrino flavor oscillations: wave packet approach}

\begin{figure}[h]
\centering
\includegraphics*[width=8.8cm]{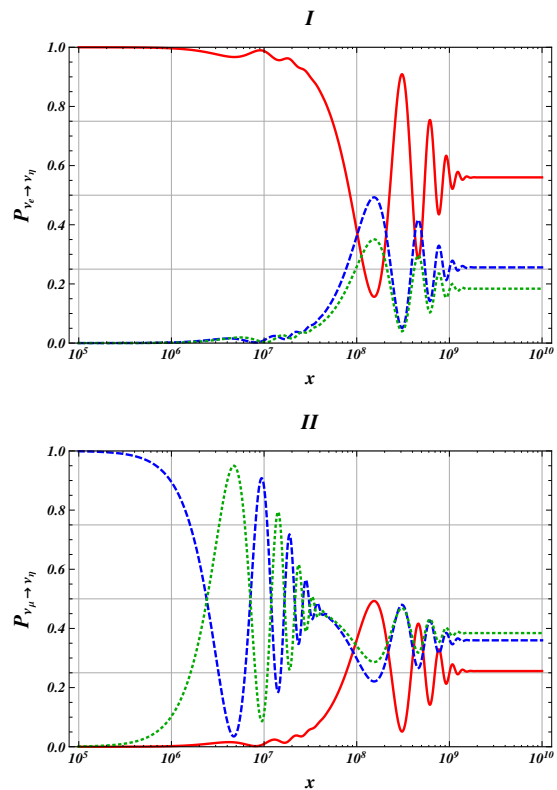}
\caption{(Color online) The transition probabilities
$P_{\nu_\alpha\longrightarrow\nu_e}(x)$ (full line),
$P_{\nu_\alpha\longrightarrow\nu_\mu}(x)$ (dashed line),
$P_{\nu_\alpha\longrightarrow\nu_\tau}(x)$ (dotted line)
as functions of the distance $x$, with $\alpha=e$ in panel I and $\alpha=\mu$ in panel II.
The CP-violating phase $\delta$ is fixed as $\delta = 0$.
The $x$ axis is in logarithmic scale, and the dimensions are meters.
The transition probabilities $P_{\nu_\tau\longrightarrow\nu_\eta}(x)$ are not reported;
for the chosen experimental parameters,
they slightly differ from the corresponding $P_{\nu_\mu\longrightarrow\nu_\eta}(x)$.
}
\label{figTransProb}
\end{figure}

The standard theory of neutrino oscillations, which describes the free evolution dynamics,
is developed using the plane-wave approximation \cite{neutroscillplanewave}.
Of course, such an approximation cuts off all the effects due to localization.
In order to recover a more realistic description of the phenomenon,
one has to resort to the wave packet approach~\cite{Nussinov,GiuntiKim,Giunti2,Giunti:2008cf}
(for reviews see Refs.~\cite{Beuthe,Giunti:2007ry}).
Following the procedure reported in Refs.~\cite{GiuntiKim,Giunti2,Giunti:2008cf},
a neutrino with definite flavor, which propagates along the $x$ direction, can be described by the state:
{
\begin{equation}
|\nu_{\alpha}(x,t)\rangle \,=\, \sum_{j} U_{\alpha j}^* \, \psi_{j}(x,t) \,|\nu_{j}\rangle \,,
\label{neutwvpack}
\end{equation}}
where $|\nu_{j}\rangle$ is the mass eigenstate of mass $m_{j}$, and $\psi_{j}(x,t)$ is its wave function;
the $U_{\alpha j}$ denotes the corresponding element of the PMNS mixing matrix in the standard form $U(\tilde{\theta},\delta)$,
with $(\tilde{\theta},\delta)\equiv (\theta_{12},\theta_{13},\theta_{23};\delta)$, being $\theta_{ij}$ the mixing angles
and $\delta$ the CP-violating phase (see {\it Eq.~(1)} of Ref.~\cite{NoiEPL}).
Assuming a Gaussian distribution $\psi_{j}(p)$ for the momentum of the massive neutrino $|\nu_{j}\rangle$
\begin{equation}
\psi_{j}(x,t) \,=\, \frac{1}{\sqrt{2\pi}} \,
\int \,dp \, \psi_{j}(p) \, e^{i p x -i E_{j}(p) t} \,,
\label{wavfunc1}
\end{equation}
the wave function writes:
\begin{equation}
\psi_{j}(p) \,=\, \frac{1}{(2\pi \sigma_{p}^{2})^{1/4}} \,
e^{-\frac{1}{4\sigma_{p}^{2}}(p-p_{j})^{2}} \,,
\label{wavfunc2}
\end{equation}
where $p_{j}$ is the average momentum, $\sigma_{p}$ is
the momentum uncertainty, and $E_{j}(p) \,=\, \sqrt{p^{2}+m_{j}^{2}}$.
The density matrix associated with the pure state, Eq.~(\ref{neutwvpack}), is given by:
\begin{equation}
\rho_{\alpha}(x,t) \,=\, |\nu_{\alpha}(x,t)\rangle \langle \nu_{\alpha}(x,t)| \,.
\label{densmatwvpack}
\end{equation}
By assuming the condition $\sigma_{p}\ll E_{j}^{2}(p_{j})/m_{j}$,
the energy $E_{j}(p)$ can be approximated by $E_{j}(p)\simeq E_{j}+v_{j}(p-p_{j})$,
with $E_{j}\equiv \sqrt{p_{j}^{2}+m_{j}^{2}}$, and
$v_{j}\equiv \frac{\partial E_{j}(p)}{\partial p}\big|_{p=p_{j}} \,=\,\frac{p_{j}}{E_{j}} $
is the group velocity of the wave packet of the massive neutrino $|\nu_{j}\rangle$.
With such an approximation, the Gaussian integration over $p$ in Eq.~(\ref{wavfunc1}) can be easily performed
(see Ref. \cite{Giunti2} for details).
In the instance of extremely relativistic neutrinos, one can exploit the following further approximations:
\begin{equation}
E_{j} \,\simeq \, E  , \quad
p_{j} \,\simeq \, E-\frac{m_{j}^{2}}{2E} , \quad
v_{j} \,\simeq \, 1-\frac{m_{j}^{2}}{2E_{j}^{2}}
\label{relapprox}
\end{equation}
where $E$ is the neutrino energy in the limit of zero mass \cite{GiuntiKim,Giunti2}.
The resulting density matrix provides a space-time description of neutrino's dynamics.
Due to the long time exposure of the detectors,
it is convenient to consider an average in time of $\rho_{\alpha}(x,t)$,
i.e. a further Gaussian integration over the time, yielding the final density matrix \cite{Giunti2}:
\begin{eqnarray}
&&\rho_{\alpha}(x) \,=\, \sum_{k,j} U_{\alpha k} U_{\alpha j}^{*}
   \; f_{jk}(x) \;
|\nu_{j}\rangle \langle\nu_{k}| \,,
\label{statwavepack}
\\
&& f_{jk}(x) \equiv \exp\left[ -i \frac{\Delta m_{jk}^{2} x}{2E}
-\left( \frac{\Delta m_{jk}^{2} x}{4\sqrt{2}E^{2}\sigma_{x}}\right)^{2}\right] ,
\label{statwavepack2}
\end{eqnarray}
where $\sigma_{x} \,=\, (2\sigma_{p})^{-1}$, and $\Delta m_{jk}^{2} \,=\, m_{j}^{2}-m_{k}^{2}$.
The parameters in Eq.~(\ref{statwavepack}), i.e. the mixing angles $\theta_{ij}$
and the squared mass differences $\Delta m_{jk}^{2}$,
and the parameters $E$ and $\sigma_{p}$, are fixed to the experimental values (see Ref.~\cite{Fogli:2006yq}):
\begin{eqnarray}
&&\sin^{2}\theta_{12} = 0.314 \;,\; \sin^{2}\theta_{13} =0.8 \times 10^{-2}\;, \nonumber
\\
&&\sin^{2}\theta_{23} =  0.45 \;, \nonumber \\
&&\Delta m_{21}^{2} = \delta m^{2} = 7.92 \times 10^{-5} \, eV^{2} \;, \\
&&\Delta m_{31}^{2} = \Delta m^{2} + \frac{\delta m^{2}}{2} \,, \qquad
\Delta m_{32}^{2} \,=\, \Delta m^{2} - \frac{\delta m^{2}}{2} \,, \nonumber  \\
&&
\Delta m^{2} \,=\, 2.6 \times 10^{-3} \, eV^{2} \;, E = 10 \,GeV \;, \sigma_{p} = 1 \,GeV \nonumber \,.
\label{sqmassdiffpar}
\end{eqnarray}
Quantum entanglement is a physical quantity that depends on the chosen observables,
and that is endowed with an operational meaning determined by the selected observables and subsystems.
In Refs.~\cite{NoiPRD,EPLBrazil}, by establishing the identification
$|\nu_{i}\rangle \,=\, |\delta_{i1}\rangle_{1}|\delta_{i2}\rangle_{2}|\delta_{i3}\rangle_{3}
\equiv |\delta_{i1}\delta_{i2}\delta_{i3}\rangle$ $(i=1,2,3)$,
the coherence of the quantum superposition of the neutrino mass eigenstates
has been investigated in terms of the spatial behavior of the multipartite entanglement
of the state (\ref{statwavepack}).
However, from an experimental point of view, it would be preferable to consider the three-qubit Hilbert space
associated with the three flavors, i.e. through the alternative identification
$|\nu_{\alpha}\rangle \,=\, |\delta_{\alpha e}\rangle_{e}|\delta_{\alpha \mu}\rangle_{\mu}|\delta_{\alpha \tau}\rangle_{\tau}
\equiv |\delta_{\alpha e}\delta_{\alpha \mu}\delta_{\alpha \tau}\rangle$ $(\alpha \,=\, e,\mu,\tau)$.
To this aim, by using the relation {$|\nu_i\rangle \,=\, \sum_{\alpha=e,\mu,\tau} \, U_{\alpha i} \, |\nu_{\alpha}\rangle$},
we rewrite the $\rho_{\alpha}(x)$ in the form:{
\begin{eqnarray}
&&\rho_{\alpha}(x) \,=\, \sum_{\beta,\gamma} \,F_{\beta \gamma}^{(\alpha)}(x) \,
|\delta_{\beta e}\delta_{\beta \mu}\delta_{\beta \tau}\rangle \langle \delta_{\gamma e}\delta_{\gamma \mu}\delta_{\gamma \tau}| \,,
\label{rhoxfinal}
\\
&&{F_{\beta \gamma}^{(\alpha)}(x)} \equiv \sum_{k,j} \; U_{\alpha j}^* U_{\alpha k}\, f_{jk}(x) \, U_{\beta j}
U_{\gamma k}^{*} \,,
\end{eqnarray}}
with $k,j=1,2,3$ and $\beta,\gamma=e,\mu,\tau$.
The transition probability for the neutrino state $\rho_{\alpha}(x)$ to be in the flavor $\eta$ at position $x$
is given by:
\begin{equation}
P_{\nu_\alpha\longrightarrow\nu_\eta}(x) \,=\, Tr[\langle \nu_\eta| \rho_{\alpha}(x) |\nu_\eta\rangle] \,=\, {F_{\eta\eta}^{(\alpha)}(x)} \,.
\label{TransProb}
\end{equation}
In Fig.~\ref{figTransProb}, we plot $P_{\nu_\alpha\longrightarrow\nu_\eta}(x)$ (for $\alpha=e,\,\mu$) as a function of $x$.
We observe that the transition probabilities are all characterized by a common feature:
if their behavior is analyzed with respect to the spatial dimension,
after an initial more stable trend they undergo an intermediate phase of wide and rapidly decreasing oscillations,
until achieving a final stationary value.
From definition (\ref{TransProb}), we see that the transition probabilities provide only
partial information; in fact, these probabilities are given by the diagonal elements $F_{\eta\eta}^{(\alpha)}$.
They do not provide any information about quantum correlations between two generic flavors
or between two generic subsystems, which are encoded in the off-diagonal elements $F_{\eta\eta'}^{(\alpha)}$,
$\eta\neq\eta'$, $\eta,\eta'=e,\mu,\tau$.
Therefore we resort to quantum information tools to gain additional insights.


\section{Entanglement measures}

The density matrix written in the form (\ref{rhoxfinal}) represents, in general, an entangled
mixed state, whose entanglement content can be quantified by means of properly devised measures.
Thus, we provide a brief recall of the definitions of the entanglement measures used in the present analysis
(for recent and detailed reviews on the qualification, quantification, and applications of entanglement,
see Refs. \cite{EntRevFazio,EntRevHorodecki}).
In the instance of mixed states, the entropic measures, such as the von Neumann entropy and the linear entropy,
cannot be used.
In order to characterize the bipartite entanglement of mixed states, several entanglement measures have been proposed, see e.g. Refs. \cite{EntFormDistill,EntRelEntr,Negativity,CoffKundWoot}. In this work, we are concerned with three-partite states, that is three flavor (three-qubit) states. In order to characterize the bipartite entanglement of such multipartite mixed states we will use two computable entanglement monotones: the concurrence \cite{CoffKundWoot} and the logarithmic negativity \cite{Negativity}. Conceptually, the logarithmic negativity is of particular interest because it has been showed that it is a full entanglement monotone notwithstanding the fact that it is not convex \cite{Plenio}.  

We anticipate that, in order to investigate the entire structure of multipartite entanglement in three-flavor mixed states, in future work \cite{noifuture}, we will study the behavior of the genuine multipartite conccurence and of the three-tangle, which were introduced and discussed in Refs. \cite{Ma,Eltschka1,Eltschka2}.

The first measure is closely related to the entanglement of formation which is the minimal amount of entanglement
needed for the production of a mixed state described by a given density matrix.
We denote by $\rho$ the density operator, corresponding to an arbitrary $N$-qubit state,
that describes a system partitioned into $N$ parties.
The reduced density operator $\rho^{(\alpha,\,\beta)}$ associated with the $\rho$ is defined as:
\begin{equation}
 \rho^{(\alpha,\,\beta)} \,=\, Tr_{\gamma\neq\alpha,\beta}[\rho] \,,
 \label{mixedrho}
\end{equation}
where the trace operation is made over all the parties different from $\alpha$ and $\beta$.
The spin-flipped state $\tilde{\rho}^{(\alpha,\,\beta)}$ reads:
\begin{equation}
\tilde{\rho}^{(\alpha,\,\beta)} = (\sigma_y\otimes\sigma_y)\rho^{(\alpha,\,\beta)*}(\sigma_y\otimes\sigma_y),
\end{equation}
where the complex conjugate is taken in the standard basis
$\left\{|00\rangle, |01\rangle, |10\rangle, |11\rangle \right\}$.
Then the concurrence is given by:
\begin{equation}
C(\rho^{(\alpha,\,\beta)})=\max\{0,\lambda_1-\lambda_2-\lambda_3-\lambda_4\},
\label{Concurrence}
\end{equation}
where $\{\lambda_i\}_{i=1}^4$ are the square roots
of the four eigenvalues of the non-Hermitian matrix $\rho^{(\alpha,\,\beta)}\tilde{\rho}^{(\alpha,\,\beta)}$,
and are non-negative real numbers taken in decreasing order with respect to the index $i$.
Therefore the concurrence can be used to measure the entanglement between two flavors in the neutrino three flavor state,
after tracing the third flavor.
We thus compute the quantities $C\left(\rho_\alpha^{(\beta,\gamma)} (x)\right) \equiv C_\alpha^{(\beta,\gamma)}$
where $\rho_\alpha^{(\beta,\gamma)} (x)= Tr_{\eta\neq\beta,\gamma}\left[\rho_\alpha (x) \right]$,
and with $\alpha,\beta,\gamma,\eta=e,\mu,\tau$.

Since the state $\rho_\alpha (x)$ possesses multipartite components,
in order to obtain a complete characterization of the entanglement content we exploit
a measure belonging to the typology of global measure of entanglement
\cite{Wallach,Brennen,Scott,Oliveira,Pascazio}.
The global measure approach relies on the construction of the set of all possible bipartitions of the total system,
that are able to encompass both bipartite and multipartite contributions.
We define a global entanglement measure for mixed states based on the logarithmic negativity
as proper measure for each bipartition.
Let $\rho$ be a multipartite mixed state associated with a system $S$,
partitioned into $N$ parties. Again, we consider the bipartition of the
$N$-partite system $S$ into two subsystems $S_{A_{n}}$ and $S_{B_{N-n}}$.
We denote by
\begin{equation}
\tilde{\rho}_{A_{n}} \equiv  \rho^{PT\, B_{N-n} }
\,=\,  \rho^{PT\, j_{1},j_{2},\ldots,j_{N-n} }
\label{rhoPT}
\end{equation}
the {\it bona fide} density matrix, obtained by the partial transposition of $\rho$
with respect to the parties belonging to the subsystem $S_{B_{N-n}}$.
The logarithmic negativity associated with the fixed bipartition will be given by
{
\begin{equation}
E_{\mathcal{N}}^{(A_{n};B_{N-n})} \,=\,
\log_{2} \parallel \tilde{\rho}_{A_{n}} \parallel_{1} \,,
\label{lognegAn}
\end{equation}}
where $\parallel \cdot \parallel_1$ denotes the trace norm.
Finally, we define the average logarithmic negativity
\begin{equation}
\langle E_{\mathcal{N}}^{(n:N-n)} \rangle \,=\, \left(%
\begin{array}{c}
  N \\
  n \\
\end{array}%
\right)^{-1} \; \sum_{A_{n}} E_{\mathcal{N}}^{(A_{n};B_{N-n})} \,,
\label{avnlogneg}
\end{equation}
where the sum is intended over all the possible bipartitions of the system.
Of course we can easily construct from Eq.~(\ref{statwavepack}) the matrix with elements
$\langle lmn|\rho_{\alpha}(x)|ijk \rangle$, where $i,j,k,l,m,n \,=\, 0,1$,
and analytically compute the quantities $E_{\mathcal{N}\,\alpha}^{(\beta,\gamma;\eta)}$,
for $\beta,\gamma,\eta=e,\mu,\tau$ and $\beta\neq \gamma\neq \eta$, and the average logarithmic negativity
$\langle E_{\mathcal{N}\,\alpha}^{(2:1)} \rangle$, for the neutrino state $\rho_\alpha (x)$
with flavor $\alpha$.

\section{Results}

\begin{figure}[h]
\centering
\includegraphics*[width=8.8cm]{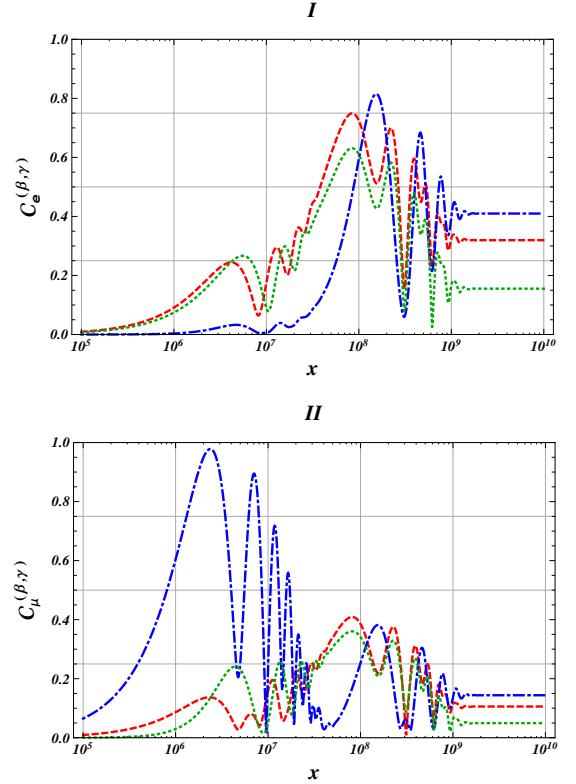}
\caption{(Color online) The concurrences $C_{\alpha}^{(\beta;\gamma)}$,
with $\alpha=e$ in panel I and $\alpha=\mu$ in panel II,
for all possible bipartitions as functions of the distance $x$.
In panel I, $C_{e}^{(e;\mu)}$ (dashed line) and $C_{e}^{(e;\tau)}$ (dotted line)
exhibit a similar behavior;
$C_{e}^{(\mu;\tau)}$ (dot-dashed line) maintain initially very low before assuming an oscillating trend.
In panel II, $C_{\mu}^{(\mu;\tau)}$ (dot-dashed) contains most of the entanglement content,
with an oscillatory and decreasing behavior;
$C_{\mu}^{(e;\mu)}$ (dashed line) and $C_{\mu}^{(e;\tau)}$ behave similarly.
The CP-violating phase $\delta$ is fixed as $\delta = 0$.
The $x$ axis is in logarithmic scale, and the dimensions are meters.
}
\label{figConca}
\end{figure}

\begin{figure}[h]
\centering
\includegraphics*[width=8.8cm]{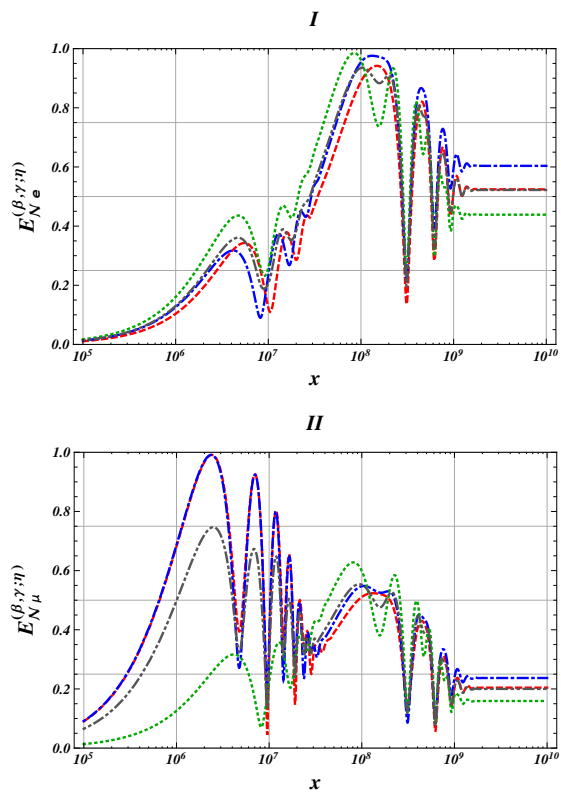}
\caption{(Color online) The logarithmic negativities $E_{\mathcal{N}\,\alpha}^{(\beta,\gamma;\eta)}$
for all possible bipartitions and the average logarithmic negativity
$\lan E_{\mathcal{N}\,\alpha}^{(2:1)} \ran$ as functions of the distance $x$.
The quantities $E_{\mathcal{N}\,e}^{(e,\mu;\tau)}$ (dashed line),
$E_{\mathcal{N}\,e}^{(e,\tau;\mu)}$ (dot-dashed line),
and $E_{\mathcal{N}\,e}^{(\mu,\tau;e)}$ (dotted line), see panel I,
exhibit a similar behavior; first after a slow increase, they show oscillations,
rapidly damped till a constant value.
In panel II the quantities $E_{\mathcal{N}\,\mu}^{(e,\mu;\tau)}$ (dashed line) and
$E_{\mathcal{N}\,\mu}^{(e,\tau;\mu)}$ (dot-dashed line) are initially quite superimposed,
they they undergo damped oscillations;
$E_{\mathcal{N}\,\mu}^{(\mu,\tau;e)}$ (dotted line) shows an initially very different behavior.
The average logarithmic negativity $\lan E_{\mathcal{N}\,\alpha}^{(2:1)} \ran$ (double-dot-dashed line)
is also reported in both panels.
The CP-violating phase $\delta$ is fixed as $\delta = 0$.
The $x$ axis is in logarithmic scale, and the dimensions are meters.
}
\label{figLogNega}
\end{figure}

We analytically compute both the concurrence $C_{\alpha}^{(\beta,\gamma)}$
and the logarithmic negativity $E_{\mathcal{N}\,\alpha}^{(\beta,\gamma;\eta)}$
associated with the state $\rho_\alpha(x)$:
{
\begin{eqnarray}
&&\hspace{-.8cm} C_{\alpha}^{(\beta,\gamma)}(x) \,=\, 2\,|F_{\beta\gamma}^{(\alpha)}(x)| \,,
\label{Concurr} \\
&&\hspace{-.8cm} E_{\mathcal{N}\,\alpha}^{(\beta,\gamma;\eta)}(x) \,=\,
\log_2 \left[ 1+2\sqrt{|F_{\beta\eta}^{(\alpha)}(x)|^2+|F_{\gamma\eta}^{(\alpha)}(x)|^2} \right]
\label{LogNegat}
\end{eqnarray}}
We observe that the above quantities are expressed in terms of off-diagonal terms.
Fig.~\ref{figConca} contains the plots of the concurrence $C_{\alpha}^{(\beta,\gamma)}$
for $\alpha=e,\,\mu$.
Looking at the curves in Fig.~\ref{figConca}, we can investigate the distribution of the entanglement
between two specific flavors. For instance in the panel I of the same figure,
the entanglement is initially distributed between the couples $(\nu_e,\nu_\mu)$ and $(\nu_e,\nu_\tau)$, while,
for greater distance, also the component $(\nu_\mu,\nu_\tau)$ acquires a growing, and then oscillating, weight.
On the contrary, in the panel II, $(\nu_\mu,\nu_\tau)$ is initially associated with the greatest component;
subsequently the entanglement is distributed quite evenly among all the components,
exhibiting an oscillatory behavior till a stabilization at a final constant value.
Nevertheless, being based on the trace performed on one flavor,
the concurrence cannot provide a complete description of the entanglement distribution.
Therefore, we exploit the global entanglement measure, built of the logarithmic negativities.
In Fig.~\ref{figLogNega} it is plotted the logarithmic negativity
$E_{\mathcal{N}\,\alpha}^{(\beta,\gamma;\eta)}$ for $\alpha=e,\,\mu$.
The curves in Fig.~\ref{figLogNega} allow to guess some indications about the distribution
of entanglement between the three bipartitions $(\beta,\gamma;\eta)$.
For example, in panel I of the figure, we can observe that all the bipartitions
possess a similar amount of entanglement as $x$ varies.
Instead in panel II, the bipartitions $(e,\mu;\tau)$ and $(e,\tau;\mu)$
exhibit a quite identical oscillatory behavior, with most of the entanglement content of the state.
The set of measures plotted in Figs.~\ref{figConca} and \ref{figLogNega}
can be exploited in a complementary way to get a clear picture of the entanglement concentration
between the flavors $(\beta,\gamma)$ and the bipartitions $(\beta,\gamma;\eta)$.
In fact, for certain regions of $x$, the entanglement is equally distributed both in the flavors and the bipartitions.
As an example, looking at the space interval $[10^8,\,10^9]$ in panel I of Fig. \ref{figConca} and
in panel I of Fig. \ref{figLogNega},
we observe that both the concurrences and the logarithmic negativities exhibit,
for some limited regions, high and comparable levels.
This evidence can be considered a signature of tripartite entanglement.

A very important aspect is that, by comparing Fig. \ref{figTransProb}
with Figs. \ref{figConca} and \ref{figLogNega}, the behavior of entanglement measures
is evidently not monotone with respect to that of transition probabilities.
This implies that the off-diagonal correlations add further insight to the understanding of the evolution of flavor neutrino states.

We see that, analogously to the transition probabilities, both $C_{\alpha}^{(\beta;\gamma)}$ and $E_{\mathcal{N}\,\alpha}^{(\beta,\gamma;\eta)}$
tend to constant values for sufficiently high distances $x$.
This fact is due to the spatial separation of the wave packets;
indeed, as a consequence, the interference effects are destroyed by the decoherence due to localization
(i.e. damping of the oscillations).
In this regime, the only surviving entanglement is the static intrinsic one \cite{NoiPRD},
due to the peculiarity of systems exhibiting particle mixing.

\section{Conclusions}
In order to gain further insight in the physics of the evolution of neutrino oscillations, we have
investigated the flavor entanglement content, and the corresponding quantum correlations when the neutrino state
is  described  by  a  time-averaged  wave-packet.
The description of the neutrino state is carried out in its natural and directly observable playground,
i.e. the flavor Hilbert space.  The flavor entanglement provides further information about quantum correlations
between different flavors and between different bipartitions of the three-flavor system.
The main aspects concerning the entanglement shared by flavors and/or flavor subsystems have
been analyzed in terms of space distribution.  In particular, the entanglement distribution,
i.e. the entanglement shared by flavors and/or flavor bipartitions, shows a peculiar behavior as a function of distance,
which cannot be inferred by looking only at transition probabilities.
It is worth to remark that an uniform distribution of entanglement among the three parties is observed in some regions of the space variable $x$. The aim of such investigations is twofold: to provide a deeper understanding of fundamental phenomena, and to lay the theoretical basis for possible practical implementations of quantum information protocols in the framework of elementary particle physics and quantum field theory.



\begin{thebibliography}{0}


\bibitem{Bert1}
  \Name{Bertlmann R.A. \and Grimus W.}
  \REVIEW{Phys. Lett. B}{392}{1997}{426}.

\bibitem{Bert2}
  \Name{Bertlmann R.A., Grimus W. \and Hiesmayr B.C.}
  \REVIEW{Phys. Lett. A}{289}{2001}{21}.

\bibitem{DiDomenico:2011fq}
 \Name{Di Domenico A., Gabriel A., Hiesmayr B.C., Hipp F., Huber M., Krizek G., M\"{u}hlbacher K., Radic S., Spengler Ch. \and Theussl L.}
 \REVIEW{Found. Phys.}{42}{2012}{778}.

\bibitem{NoiPRD}
 \Name{Blasone M., Dell'Anno F., De Siena S., Di Mauro M. \and Illuminati F.}
 \REVIEW{Phys. Rev. D}{77}{2008}{096002}.

\bibitem{NoiEPL}
 \Name{Blasone M., Dell'Anno F., De Siena S. \and Illuminati F.}
 \REVIEW{Europhys. Lett.}{85}{2009}{50002}.

\bibitem{Blasone:2013yoa}
 \Name{Blasone M., Dell'Anno F., De Siena S. \and Illuminati F.}
 \REVIEW{J. Phys. Conf. Ser.}{442}{2013}{012070}.

\bibitem{Kayser10}
 \Name{Kayser B., Kopp J., Hamish Robertson R.G. \and Vogel P.}
 \REVIEW{Phys. Rev. D}{82}{2010}{093003};

\bibitem{Smirnov11}
 \Name{Akhmedov E.K. \and Smirnov A.Y.}
 \REVIEW{Found. Phys.}{41}{2011}{1279};


\bibitem{Boyanovsky:2011xq}
 \Name{Boyanovsky D.}
 \REVIEW{Phys. Rev. D}{84}{2011}{065001}

\bibitem{Lello:2013bva}
 \Name{Lello L., Boyanovsky D. \and Holman R.}
 \REVIEW{JHEP}{11}{2013}{116}.

\bibitem{NoiEPL2}
\Name{Blasone M., Dell'Anno F., De Siena S. \and Illuminati F.}
\REVIEW{Europhys. Lett.}{106}{2014}{30002}.

\bibitem{NoiHindawi}
\Name{Blasone M., Dell'Anno F., De Siena S. \and Illuminati F.}
\REVIEW{Adv. High Energy Phys.}{2014}{2014}{359168}.

\bibitem{EPLBrazil}
\Name{Bittencourt V.A.S.V., Villas Boas C.J. \and Bernardini A.E.}
\REVIEW{Europhys. Lett.}{108}{2014}{50005}.

{
\bibitem{Bernardini:2012uf}
 \Name{Bernardini, A.E. \and Bittencourt, V.A.S.V.}
 \REVIEW{Astropart. Phys.}{41}{2013}{31}.


\bibitem{Alok:2014gya}
\Name{Alok A.~K., Banerjee S. \and Sankar S.~U.},
arXiv:1411.5536 [quant-ph]

\bibitem{Banerjee}
\Name{Banerjee S., Alok A.~K.,Srikanth R.~ \and Hiesmayr B.~C.},
arXiv:1508.03480[hep-ph]

\bibitem{neutroscillplanewave}
  \Name{Bilenky S.M. \and Pontecorvo B.}
  \REVIEW{Phys. Rep.}{41}{1978}{225}.

\bibitem{Nussinov}
  \Name{Nussinov S.}
  \REVIEW{Phys. Lett. B}{63}{1976}{201}.

\bibitem{GiuntiKim}
  \Name{Giunti C. \and Kim C.W.}
  \REVIEW{Phys. Rev. D}{58}{1998}{017301}.

\bibitem{Giunti2}
  \Name{Giunti C.}
  \REVIEW{Found. Phys. Lett.}{17}{2004}{103}.

\bibitem{Giunti:2008cf}
  \Name{Giunti C.}
  \REVIEW{J. Phys. G Nucl. Part. Phys.}{34}{2007}{93}.

\bibitem{Beuthe}
  \Name{Beuthe M.}
  \REVIEW{Phys. Rep.}{375}{2003}{105}.
}

\bibitem{Giunti:2007ry}
\Name{Giunti C. \and Kim C.W.}
 \Book{Fundamentals of Neutrino Physics and Astrophysics}
\Publ{Oxford University Press, Oxford}
\Year{2007}.

\bibitem{Fogli:2006yq}
 \Name{Fogli G.L., Lisi E., Marrone A., Melchiorri A., Palazzo A., Serra P., Silk J. \and Slosar A.}
 \REVIEW{Phys. Rev. D}{75}{2007}{053001}.




\bibitem{Marinatto}
\Name{Ghirardi G.C., Marinatto L. \and Weber T.}
\REVIEW{J. Stat Phys.}{108}{2002}{49};

\bibitem{Benatti}
\Name{Benatti F., Floreanini R. \and Marzolino U.}
\REVIEW{Ann. Phys.}{327}{2012}{1304};
\REVIEW{Ann. Phys.}{325}{2010}{924}.

\bibitem{Zanardi}
\Name{Zanardi P.}
\REVIEW{Phys. Rev. A}{65}{2002}{042101};
%
\Name{Zanardi P., Lidar D.A. \and Lloyd S.}
\REVIEW{Phys. Rev. Lett.}{92}{2004}{060402}.

\bibitem{Wiseman}
\Name{Wiseman H.M. \and Vaccaro J.A.}
\REVIEW{Phys. Rev. Lett.}{91}{2003}{097902};

\bibitem{Viola}
\Name{Barnum H. et al.}
\REVIEW{Phys. Rev. Lett.}{92}{2004}{107902};

\bibitem{Cirac}
\Name{Banuls M.-C., Cirac J.I. \and Wolf M.M.}
\REVIEW{Phys. Rev. A}{76}{2006}{022311}.

\bibitem{VanEnk}
\Name{van Enk S.J.}
\REVIEW{Phys. Rev. A}{72}{2005}{064306}.

\bibitem{Vedral}
\Name{Terra Cunha M.O., Dunningham J.A. \and Vedral V.}
\REVIEW{Proc. R. Soc. A}{463}{2007}{2277}.

%
%

%
%


%








\bibitem{EntRevFazio}
 \Name{Amico L., Fazio R., Osterloh A. \and Vedral V.}
 \REVIEW{Rev. Mod. Phys.}{80}{2008}{517}.

\bibitem{EntRevHorodecki}
\Name{Horodecki R., Horodecki P., Horodecki M. \and Horodecki K.}
\REVIEW{Rev. Mod. Phys.}{81}{2009}{865}.

\bibitem{EntFormDistill}
 \Name{Bennett C.H., Di Vincenzo D.P., Smolin J.A. \and Wootters W.K.}
 \REVIEW{Phys. Rev. A}{54}{1996}{3824}.

\bibitem{EntRelEntr}
 \Name{Vedral V. \and Plenio M.B.}
 \REVIEW{Phys. Rev. A}{57}{1998}{1619}.

\bibitem{Negativity}
 \Name{Vidal G. \and Werner R.F.}
 \REVIEW{Phys. Rev. A}{65}{2002}{032314}.

\bibitem{CoffKundWoot}
 \Name{Coffman V., Kundu J. \and Wootters W.K.}
 \REVIEW{Phys. Rev. A}{61}{2000}{052306}.

\bibitem{Plenio}
\Name{Plenio M.B.}
\REVIEW{Phys. Rev. Lett.}{95}{2005}{090503}.

\bibitem{noifuture}
\Name{Blasone M., Dell'Anno F., De Siena S. \and Illuminati F.}
\REVIEW{in preparation}{}{2016}{}. 

\bibitem{Ma}
\Name{Ma Z.-H., Chen Z.-H., Chen J.-L., Spengler C., Gabriel A. \and Huber M.}
\REVIEW{Phys. Rev. A}{83}{2011}{062325}.

\bibitem{Eltschka1}
\Name{Eltschka C \and Siewert J.}
\REVIEW{Phys. Rev. A}{89}{2014}{022312}.

\bibitem{Eltschka2}
\Name{Eltschka C \and Siewert J.}
\REVIEW{J. Phys. A: Math. Theor.}{47}{2014}{424005}.



\bibitem{Wallach}
\Name{Meyer D.A. \and Wallach N.R.}
\REVIEW{J. Math. Phys.}{43}{2002}{4273}.

\bibitem{Brennen}
\Name{Brennen G.K.}
\REVIEW{Quantum Inf. Comp.}{3}{2003}{619}.


\bibitem{Scott} A. J. Scott, Phys. Rev. A {\bf 69}, 052330 (2004).

\bibitem{Oliveira}
\Name{de Oliveira T.R., Rigolin G. \and de Oliveira M.C.}
\REVIEW{Phys. Rev. A}{73}{2006}{010305(R)}.

\bibitem{Pascazio}
 \Name{Facchi P., Florio G. \and Pascazio S.}
 \REVIEW{Phys. Rev. A}{74}{2006}{042331}.









%
%
%









\end{thebibliography}
\end{document}